\documentclass[prd,twocolumn,aps,amsmath,amssymb,nofootinbib,preprintnumbers]{revtex4}

\usepackage{amsmath,amssymb,epsf,graphicx}

\def\lesssim{\mathrel{\hbox{\rlap{\hbox{\lower4pt\hbox{$\sim$}}}\hbox{$<$}}}}
\def\gtrsim{\mathrel{\hbox{\rlap{\hbox{\lower4pt\hbox{$\sim$}}}\hbox{$>$}}}}
\def\Mp{M_{\mathrm{Pl}}}
\def\be{\begin{equation}}
\def\ee{\end{equation}}
\def\bea{\begin{eqnarray}}
\def\eea{\end{eqnarray}}

\begin{document}
\title{Gravitational Waves from the Non-Perturbative Decay of Condensates along Supersymmetric Flat Directions}
\date{\today}
\author{Jean-Fran\c{c}ois Dufaux}
\affiliation{Instituto de F\'isica Te\'orica \ UAM/CSIC,
Universidad Aut\'onoma de Madrid, Cantoblanco, 28049 Madrid, Spain}

\preprint{IFT-UAM/CSIC-09-07}

\begin{abstract}
It has recently been shown that specific non-perturbative effects may lead to an explosive decay of flat direction 
condensates in supersymmetric theories. We confirm explicitly the efficiency of this process with lattice 
simulations: after few rotations of the condensates in their complex plane, most of their energy
is quickly converted into inhomogeneous fluctuations. We then point out that this generates a gravitational wave background
which depends on the inflaton sector and falls in the Hz-kHz frequency range today. We compute the resulting spectrum 
and study how it depends on the parameters. We show that these gravity waves can be observable by upcoming experiments 
like Advanced LIGO and depend crucially on (i) the initial VEV of flat directions when they start to oscillate, 
(ii) their soft SUSY-breaking mass and (iii) the reheat temperature of the universe.
This signal could open a new observational window on inflation and low-energy supersymmetry.
\end{abstract}

\maketitle

\noindent

Gravity wave (GW) experiments could provide unique informations about high-energy 
phenomena in the early universe, such as inflation, preheating after inflation, cosmic strings and 
first order phase transitions \cite{review}. In particular, cosmological GW backgrounds will be a 
target for several high-sensitivity interferometric experiments, planned or proposed, in the frequency 
range between $10^{-5}$ and $10^3$ Hz. In this letter, we want to consider a new cosmological source of 
GW, which emerges naturally in the framework of high-energy physics.

Supersymmetric theories typically involve many flat directions (see \cite{mssm} for the MSSM),
i.e. directions in field space where the renormalizable part of the scalar potential is exactly 
flat in the limit of unbroken supersymmetry. 
The flatness is lifted by soft SUSY-breaking terms, non-renormalizable 
terms and SUSY-breaking terms from the finite energy density in the early universe \cite{DRT}.
Scalar field condensates may develop large VEVs along these directions during 
inflation and have several interesting consequences in cosmology \cite{EM}. A major example is the 
Affleck-Dine mechanism for baryogenesis~\cite{AD}.
Condensates with large VEVs start to oscillate when the Hubble rate $H$ becomes of the 
order of their soft mass $m \sim$ TeV \cite{C1}, with an initial amplitude that is model-dependent 
and could be as high as the Planck scale~\cite{DRT, olive}. It has recently 
been shown \cite{marco1, marco2, basboll}, that non-perturbative resonant effects 
may lead to an explosive decay of these coherent oscillations. We will show that this generates 
GW that can be observable by upcoming experiments \cite{C2}.

The resonant decay of flat directions shares many similitudes with preheating after inflation, 
when the inflaton decays into large, non-thermal fluctuations of itself and other bosonic 
fields~\cite{KLS}. GW from preheating have been intensively studied recently
\cite{pregw1, pregw2, pregw, fraginf}.
Ref.~\cite{pregw1} developed theoretical
and numerical methods to calculate GW production from dynamical scalar fields in an expanding universe.
We will apply and extend these methods to the decay of flat directions.

The resulting GW, however, will have different properties. GW from preheating have been studied so far in two 
main classes of models: parametric resonance after chaotic inflation and tachyonic preheating after hybrid 
inflation. In the first case, the typical GW frequency today is of the order of $f \sim 10^7$ Hz, which is 
too high to be observable. Hybrid inflation models may occur at lower energy scales, so it was initially 
hoped that GW from tachyonic preheating may fall into an observable range. However, as was shown in \cite{pregw2},
this generically requires very small coupling constants \cite{hyb}. By contrast, we will see that GW from the
non-perturbative decay of flat directions fall naturally in the Hz-kHz frequency range today, where
high-sensitivity interferometric experiments can operate. Another difference - which will be crucial for the
GW amplitude - is that, whereas the inflaton dominates the energy density before it decays, flat directions are
usually subdominant (at least for sub-Planckian VEVs and when their non-perturbative decay is sufficiently fast).
During their decay, the universe can then be matter-dominated if the inflaton is still oscillating around the 
(quadratic) minimum of its potential, or radiation-dominated if it decays earlier and its decay products thermalize
when $H > m$. Each case leads to different GW spectra, so GW from the non-perturbative decay of flat directions
carry also informations on the inflaton sector.

To illustrate why it may be difficult for GW from preheating to be observable, consider a simple 
model, $V = m^2 \phi^2 + g^2 \phi^2 \chi^2$, with two {\it real} scalar fields $\phi$ and $\chi$.  
In this model, $\chi$-particles are produced due to the non-adiabatic evolution of their frequency 
as the condensate $\phi$ oscillates around the minimum~\cite{KLS}. For $m \sim$ TeV, this could {\it a priori} 
describe the decay of a flat direction $\phi$. We will denote by an index "i" the time $t_i$ when the flat 
direction starts to oscillate ($H_i = m$), by $\Phi_i$ its initial amplitude at that time and we normalize 
the scale factor to $a_i = 1$. The spectra of the quanta amplified by preheating are usually strongly 
peaked around some typical (comoving) momentum $k_*$, which is inherited by the GW spectrum 
\cite{fraginf, pregw1, pregw2}. For the model above, $k_* \sim \sqrt{g m \Phi_i} a^{1/4}$ \cite{KLS}.
Assuming for simplicity that the universe is already radiation-dominated during the decay, this leads to the 
estimate $f_* \sim \sqrt{g \Phi_i / \Mp}\, 10^{11}$ Hz for the peak frequency of the resulting GW today. Taking 
for instance $\Phi = 10^{-2}\,\Mp$, $f_* < 10^3$ Hz requires an extremely small coupling constant $g^2 < 10^{-28}$.
If $\Phi_i$ is smaller, the GW amplitude is too low to be observable. Indeed, since the energy density in the 
flat direction $\rho_{\mathrm{flat}}$ is subdominant, only a small fraction
$\rho_{\mathrm{flat}} / \rho_{\mathrm{tot}}$ of the total energy density is available for the production of GW.
We will see that the fraction of energy density in GW varies as $(\rho_{\mathrm{flat}} / \rho_{\mathrm{tot}})^2$,
which can be very small.

However, flat directions are complex fields and non-renormalizable terms usually generate a velocity for their
phase when $H \sim m$ \cite{AD, DRT}. The subsequent dynamics is dominated by the $m^2 |\phi|^2$ term, resulting in 
a constant, non-zero phase velocity and an elliptical motion for $a^{3/2} \phi$ in its complex plane. This leads to time-dependent (and quasi-periodic) mixings between different excitations around flat directions, which may result
in a very efficient mechanism of particle production \cite{marco1, marco2}, significantly different from the model 
considered above. In particular, these particles have a typical momentum $k_* \sim m$, of the order of the Hubble scale. Formally, this amounts to replace $\sqrt{g \Phi_i}$ by $\sqrt{m}$ in the previous model. This gives a peak frequency of
GW that is independent of $\Phi_i$ and of order $10^3$ Hz today for $m \sim$ TeV.

This encouraging estimate should be complemented by detailed computations. The resonant decay of flat directions 
has been studied so far only in the linear regime, neglecting any backreaction of the quanta produced. Backreaction is 
eventually responsible for most of the decay and we have to use lattice simulations to fully take it into account. We 
will simulate the model \cite{marco1, CI}
\be
\label{model}
V = m^2\,|\phi|^2 + m_\chi^2\,|\chi|^2 + \frac{g^2}{2}\,\left(\phi\,\chi^* + \phi^*\,\chi\right)
\ee
where $*$ denotes the complex conjugate, $\phi$ the flat direction field and $\chi$ its decay products.
This potential includes the relevant terms for excitations around MSSM flat directions (where the interaction
comes from the D-term) but at least two flat directions have to be considered when the symmetry broken by their 
VEV is gauged \cite{marco1}. This case was studied in \cite{marco2, basboll} and found to be largely similar to 
the model (\ref{model}), so here we focus on (\ref{model}) for simplicity. We perform
lattice simulations with the parallel code ClusterEasy \cite{gary} and we compute the resulting
GW spectrum with the method of \cite{pregw2}, extended here to take into account the expansion of the universe.

\begin{figure}[hbt]
\vspace*{-0.3 cm}
\centering
\includegraphics[width=0.65\columnwidth, angle=270]{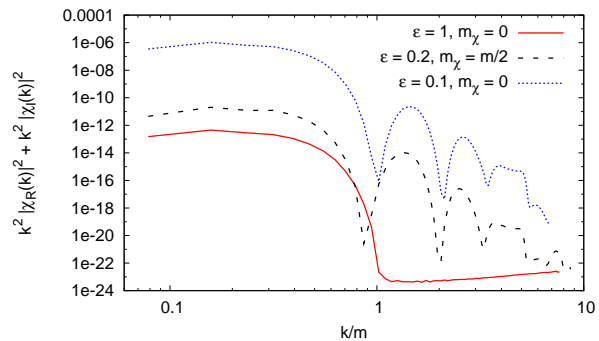}
\vspace*{-0.7 cm}
\caption{Spectra of the $\chi$ modes without expansion of the universe when backreaction is still negligible.
They are independent of $q$ as far as $q \epsilon^4 \gg 1$.
For $q \epsilon^4 \ll 1$, particle production is much less efficient.}
\label{mink}
\end{figure}

\begin{figure}[hbt]
\vspace*{-0.2 cm}
\begin{center}
\centering \leavevmode \epsfxsize=7.5cm
\epsfbox{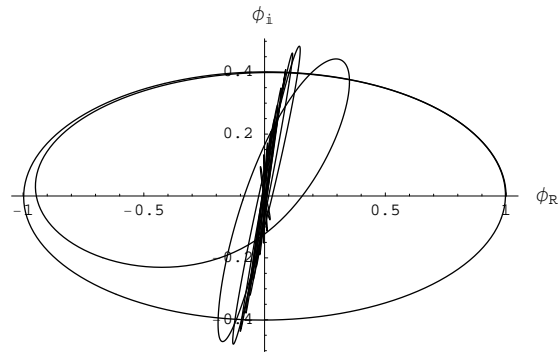}
\vspace*{-0.3 cm}
\caption{Evolution of the mean of the real and imaginary parts of $a^{3/2} \phi$ in a matter-dominated background 
with $\epsilon = 0.4$, $g^2 = 0.01$, $m = 1$ TeV and $\Phi_i = 10^{-12} \Mp$. \vspace*{-0.3 cm}}
\label{exp}
\end{center}
\end{figure}

Consider first the non-perturbative decay itself. The background evolution of the condensate 
is given by
$\phi = \left(\cos mt + i\,\epsilon\,\sin mt\right)\,\Phi_i / \sqrt{2 a^3}$. The "ellipticity"
$\epsilon \leq 1$ is model-dependent but typically of order $\epsilon \sim 0.1$ \cite{DRT}. An analytical
study of particle production in this background will appear elsewhere. Here we 
just present the results. It is convenient to introduce the parameter $q = q_i / a^3$ with $q_i = g^2 \Phi_i^2 / m^2$. 
Note that $q_i$ can be as high as $10^{30}$ for $\Phi_i$ close to the Planck scale. In the realistic case
$q \epsilon^4 \gg 1$ \cite{real}, particle production occurs because the physical eigenstates vary with time 
\cite{marco1}. Fig.~\ref{mink} shows the spectra of the $\chi$-modes 
in a Minkowski background 
when backreaction is still negligible.
The resonance is characterized by large instability bands separated by narrow stability bands, 
contrary to parametric resonance but similarly to the tachyonic resonance
studied in the second ref. of \cite{eos}.
When the trajectory of the condensate in its complex plane is circular,
$\epsilon = 1$, there is a tachyonic amplification 
of the modes with $k^2 + m_\chi^2 < m^2$ \cite{marco1}. When $\epsilon < 1$, the amplification is more efficient 
and extra instability bands appear in the UV, roughly up to $k < m / \epsilon$. The IR band has the 
highest amplitude, except if $m_\chi$ is very close to $m$. The peak of the spectrum is located at 
$\kappa_* \equiv k_* / m \sim 0.1$. In an expanding universe, 
the physical momenta $k/a$ "move" in the resonance pattern and the peak is shifted to a {\it comoving} momentum 
$\kappa_* \sim 1$. When the $\chi$-modes have been sufficiently amplified, they backreact on the condensate and 
convert its energy into inhomogeneous fluctuations of both fields. This will be responsible for most of the GW production.
We show the evolution of the mean of $a^{3/2} \phi$ in its complex plane in Fig.~\ref{exp}. The field first follows
its elliptical trajectory, but the coherent motion is then quickly destroyed by backreaction. 
This typically happens after only $1$ to $5$ rotations of the condensate (the higher $g^2$, the sooner it occurs).

We now turn to the actual computation of GW. 
Their evolution from the time of production (that we will denote with an index "p") up to now depends on the 
background equation of state, see e.g.~\cite{pregw1} for details. In our case,
their present-day frequency and energy density per logarithmic frequency interval \cite{review} read
\bea
\label{spectoday}
f &=& \frac{k}{\rho_i^{1/4}}\,\left(\frac{a_i}{a_{RD}}\right)^{1/4}\,4 \times 10^{10}\,\mathrm{Hz} 
\nonumber \\
h^2 \Omega_{\mathrm{gw}} &=& 9.3 \times 10^{-6}\,\frac{1}{\rho_i}\,
\left(\frac{a^4}{a_i^4} \frac{d \rho_{\mathrm{gw}}}{d \ln k}\right)_p \, \left(\frac{a_i}{a_{RD}}\right)
\eea
where $k$ is the comoving wave-number and $\rho_i$ is the {\it total} energy density at $t_i$.
If the universe is matter-dominated (MD) during the flat direction decay and becomes radiation-dominated (RD) 
later, when $a = a_{RD}$, then GW are further redshifted and diluted by the factors in $a_i/a_{RD}$ above.
These factors are absent if the universe is already RD at $t_i$ \cite{ard}.

We can estimate how the GW spectrum depends on the parameters
from the characteristic physical length $R_* = a/k_*$ of the scalar field inhomogeneities \cite{fraginf,pregw1,pregw2},
but we have to take into account that the flat direction energy density $\rho_{\mathrm{flat}}$ is usually subdominant.
We then estimate the fraction of energy density in GW at the time of production as \cite{fracgw}
$\rho_{\mathrm{gw}} / \rho_{\mathrm{tot}} \sim 0.1\,\left(R_* H\right)^2\,\left(\rho_{\mathrm{flat}}/\rho_{\mathrm{tot}}\right)^2$.
Using $\rho_{\mathrm{flat}\,i} = m^2 \Phi_i^2$, $k_* = \kappa_* m$ and $H_i = m$, Eqs.(\ref{spectoday}) give
\bea
\label{fstar}
f_* &\simeq& \kappa_* \, \sqrt{\frac{m}{\mathrm{TeV}}} \, \left(\frac{a_i}{a_{RD}}\right)^{1/4} \, 5 \times 10^2 \, \mathrm{Hz} \\
\label{omegastar}
h^2 \Omega_{\mathrm{gw}}^* &\sim& \frac{10^{-4}}{\kappa_*^2} \, \left(\frac{\Phi_i}{\Mp}\right)^4 \, \left(\frac{a_i}{a_{RD}}\right)
\eea
for the frequency and amplitude of the GW spectrum's peak. Note that the amplitude is very sensitive to $\Phi_i$.
Note also that Eq.(\ref{omegastar}) cannot be extrapolated to arbitrary large $\Phi_i$. If a flat direction acquires
an initial VEV of order $\Mp/3$ or larger, it dominates the total energy density before decaying and its oscillations are
delayed until $\Phi_i \sim \Mp/3$. Thus Eq.(\ref{omegastar}) with $\Phi_i = \Mp/3$ can be considered as an upper limit on 
the GW amplitude, i.e. $h^2 \Omega_{\mathrm{gw}}^* \lesssim 10^{-6}$. The same bound follows from \cite{fracgw} with
$R_* \lesssim 1 / H$ and $\rho_{\mathrm{flat}} \lesssim \rho_{\mathrm{tot}}$ \cite{pregw2}.
\begin{figure}[hbt]
\begin{center}
\vspace*{-0.3 cm}
\centering \leavevmode \epsfxsize=7.5cm
\epsfbox{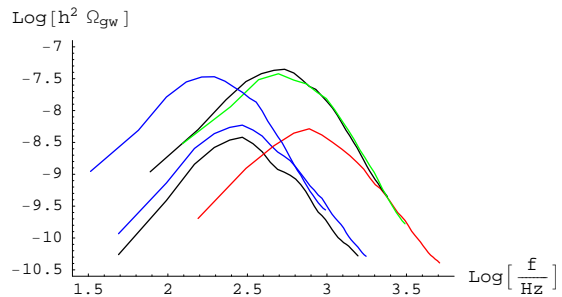}
\vspace*{-0.5 cm}
\caption{Lattice results for the GW spectra today, normalized as $h^2 \Omega_{\mathrm{gw}} \times \Mp^4 / (10 \Phi_i)^4$. 
Here we did not include the $a_i/a_{RD}$ factors in (\ref{spectoday}).
{\it Top-left}: RD background, $\Phi_i = 10^{-13} \Mp$, $m = 122$ GeV, $m_\chi = 0$, $q_i = 10^6$, $\epsilon = 0.5$ (blue).
{\it Top-right}: RD background, $\Phi_i = 10^{-1} \Mp$, $m = 1.22$ TeV, $m_\chi = 0$, (i) $q_i = 10^6$, $\epsilon = 0.5$ 
(black) and (ii) $q_i = 10^8$, $\epsilon = 0.2$ (green).
{\it Bottom-left}: MD background, $\Phi_i = 10^{-13} \Mp$, $m = 122$ GeV, $q_i = 10^6$, (i) $m_\chi = 0$, $\epsilon = 0.5$ 
(blue) and (ii) $m_\chi = m/5$, $\epsilon = 1$ (black).
{\it Bottom-right}: MD background, $\Phi_i = 10^{-1} \Mp$, $m = 1.22$ TeV, $m_\chi = 0$, $q_i = 10^6$, $\epsilon = 0.5$ (red).
\vspace*{-0.5 cm}}
\label{omega}
\end{center}
\end{figure}

We confirmed the predictions (\ref{fstar}, \ref{omegastar}) with intensive lattice simulations.
Two different scales have to be kept under control numerically: the mass and the VEV of the condensate,
whose ratio is measured by $q_i$. In our simulations, $m$ defines the natural unit of time while $1/\sqrt{q_i}$ 
defines the time step required to accurately follow the dynamics. Clearly, we cannot simulate values of $q_i$ as
high as $10^{30}$. However, we saw above that the linear stage of particle production is independent of $q_i$, see also 
\cite{marco1}. Indeed, the GW spectra that we obtained for different $q_i$ (by varying $g^2$) were nearly identical, 
as far as $q \epsilon^4 \gg 1$ was satisfied. We were able to simulate values of $q_i$ ranging from $10^4$ to $10^8$, 
which allowed us to take $\epsilon$ ranging from $0.2$ to $1$. Smaller values of $\epsilon$ lead to faster particle 
production, which could increase the GW amplitude. Increasing the mass of $\chi$ has the opposite effect and makes the 
UV modes relatively more important (which makes this case more difficult to simulate). However, this  
becomes negligible when $m_\chi$ differs significantly from $m$. This is probably more natural in models with multiple flat
directions \cite{marco2}. We varied $g^2$ from $10^{-24}$ to $0.5$, $\Phi_i$ from $10^{-13} \Mp$ to $10^{-1} \Mp$ and $m$ 
from $100$ GeV to $10$ TeV, finding excellent agreement with Eqs.~(\ref{fstar}, \ref{omegastar}), see Fig.~\ref{omega}.
Note that $\kappa_* \sim 1$ is slightly smaller in a RD background than in a MD one, leading to GW with higher amplitude
and smaller frequency at the time of production.

\begin{figure}[hbt]
\centering
\includegraphics[width=\columnwidth]{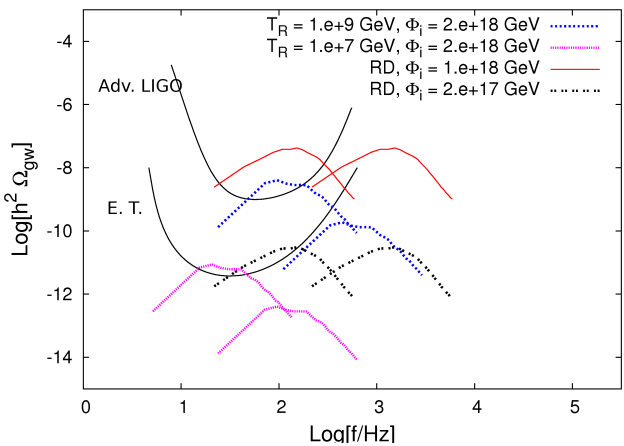}
\vspace*{-0.4 cm}
\caption{Sensitivity sketch of Advanced LIGO and Einstein Telescope \cite{review} compared to
GW spectra produced in the model (\ref{model}). In each case, the spectrum at the left is for $m = 100$ GeV and 
the one at the right for $m = 10$ TeV.}
\label{sensi}
\end{figure}

In Fig.~\ref{sensi}, we compare our results with the sensitivity of interferometric experiments.
We consider the soft masses $m = 100$ GeV and $m = 10$ TeV, different initial VEVs $\Phi_i$ and
different effective temperatures $T_R \propto \rho_{RD}^{1/4}$ when the universe becomes radiation-dominated.
If the inflaton decays slowly (perturbatively) and its decay products quickly thermalize, then $T_R$ may correspond
to the usual reheat temperature of the universe. In general, however, the universe may become radiation-dominated
long before full thermal equilibrium is established, see e.g.~\cite{eos}. In particular, this means
that the standard bound from the thermal over-production of gravitinos does not necessarily apply on $T_R$.
For $T_R \gtrsim 0.2 \sqrt{m \Mp}$, the universe is RD during the flat direction decay
and the GW amplitude is independent of $m$. If $T_R$ is lower, GW are further diluted and redshifted
by the factors in $a_i / a_{RD}$ in Eqs.(\ref{spectoday}), which depend on $m$. Assuming a fast transition
from MD to RD, we have $a_i / a_{RD} \sim (5 T_R / \sqrt{m \Mp})^{4/3}$ in that case.
We show the results for $T_R = 10^9$ GeV \cite{TR} and $10^7$ GeV. We see that
these GW can be observable for a reasonable range of soft masses and $T_R$, but this 
generically requires that some flat directions acquire a large enough initial VEV. A detection would be easier
for low $m$ or high $T_R$, but we should have at least $\Phi_i > 10^{17}$ GeV.

The initial amplitude of flat direction oscillations depends on the lowest order
of the non-renormalizable terms that lift the flatness of each direction \cite{DRT}. A superpotential term
like $\phi^n / M^{n-3}$ (with $M$ the cutoff scale) leads to
$\Phi_i \sim (m M^{n-3})^{1/(n-2)}$ \cite{DRT} when $H \sim m$. In the MSSM, $n=9$ is the minimal value allowed
by gauge invariance for the flattest direction \cite{mssm}. This gives $\Phi_i \sim 10^{16}$ GeV for $M$ the reduced 
Planck mass, which would be too small for the GW computed here to be observable by currently proposed experiments.
However, the fact that this term is allowed by gauge invariance does clearly not imply that it is indeed present,
e.g. it can be forbidden by other symmetries. Higher-order terms lead to larger values of $\Phi_i$ and therefore
to GW that can be observable. Indeed, initial VEVs can be as high as the Planck scale \cite{DRT},
e.g. in no-scale supergravity models \cite{olive}.

To summarize, we performed the first non-linear study of the non-perturbative decay of supersymmetric
flat directions with lattice simulations. This explosive process can have important consequences
in cosmology, in particular for baryogenesis and reheating after inflation. We showed that it generates
a stochastic background of GW that is very sensitive to the initial VEV of flat directions when they start
to oscillate, to their soft SUSY-breaking mass and to the time when the universe becomes radiation-dominated.
The specific spectral properties that we computed can be used to distinguish these GW from other possible
cosmological backgrounds. Of particular interest is that they fall in the Hz-kHz frequency range, where the
signal is not expected to be significantly screened by stochastic backgrounds from astrophysical sources.
We showed that, even in a simple model, these GW can be observable by upcoming experiments, including Advanced LIGO,
if some flat directions acquire a large enough initial VEV. This typically requires the flatness of the potential
to be protected up to high order by symmetries, which is model-dependent. Our results strongly motivate the study
of more complex models, with multiple flat directions and local gauge invariance \cite{marco1, marco2}. In particular,
vector fields are likely to increase the range of parameters leading to an observable signal \cite{pregw1}.
Any possibility of direct detection has to be studied in depth, because it would provide crucial informations on
both supersymmetry and the very early universe.

{\it Acknowledgments-} I thank Marco Peloso for important discussions, Juan Garcia-Bellido and Lev Kofman for
useful comments on the manuscript and Gary Felder for his work on ClusterEasy. I acknowledge the intensive use
of the IFT computation cluster and support by the Spanish MEC via FPA2006-05807 and FPA2006-05423.




\begin{thebibliography}{99}
\bibitem{C1} 

Thermal effects make condensates with moderate VEVs oscillate earlier, but not the condensates 
with high VEVs that we will be interested in (because the fields coupled to them are too massive
to be in thermal equilibrium) \cite{thermal}.
We assume that inflation occurs at $H_{\mathrm{inf}} >$ TeV and gravity-mediated SUSY breaking.
\bibitem{C2}
Refs.~\cite{qball} consider GW from the fragmentation of condensates into Q-balls. 
The non-perturbative decay that we consider occurs on significantly shorter timescales, 
so the fragmentation into Q-balls is not relevant here.
\bibitem{hyb}
Lowering the energy scale during inflation decreases the present-day frequency of GW from preheating
if their wavelengths at the time of production is a fixed fraction of the Hubble radius at that
time. However, in most models, this fraction varies with the parameters and the resulting frequency
today depends mainly on some coupling constants.
\bibitem{CI}
In a complete model, the potential includes also non-renormalizable and Hubble-dependent terms, which 
generate an initial phase velocity for the condensate when it starts to oscillate. However, these 
model-dependent terms become quickly negligible as the universe expands. We thus neglect these terms in 
the simulations and treat the initial phase velocity as a free parameter.
\bibitem{real}
The case $q\,\epsilon^4 \ll 1$ is similar to the model with two real scalar fields mentioned earlier.
\bibitem{ard}
We found that the equation of state of the flat direction decay products jumps quickly towards $w = 1/3$, 
as in \cite{eos}, so usually the universe remains RD if it is so at $t_i$.
\bibitem{TR}
This corresponds to an inflaton with a mass of order $10^{13}$ GeV and only
gravitational couplings. The usual gravitino bound is marginally satisfied in this case.
\end{thebibliography}
\end{document}